\documentclass[11pt]{article}
\usepackage{graphicx} 
\usepackage{amsmath}
\usepackage{lipsum}
\usepackage{pdfcomment}

\usepackage[table,xcdraw]{xcolor} 
\usepackage{subcaption}
\usepackage[utf8]{inputenc}
\usepackage{longtable}
\usepackage{graphicx}
\usepackage{float}
\usepackage{booktabs}
\usepackage{float} 
\usepackage{geometry}
\usepackage{lineno}
\usepackage{authblk}
\usepackage{tcolorbox}
\usepackage{rotating}
\usepackage{booktabs}
\usepackage{threeparttable}
\usepackage{geometry} 
\usepackage{lscape}
\usepackage{siunitx}  
\usepackage{adjustbox} 
\usepackage{pgfplots}
\usepackage{caption}
\usepgfplotslibrary{groupplots}

\usepackage[numbers,sort&compress]{natbib}

\usepackage{hyperref}
\hypersetup{
  colorlinks=true,
  linkcolor=blue,       
  citecolor=blue,       
  urlcolor=blue         
}

\usepackage{cite}

\usepackage{array}
\usepackage{multirow}

\usepackage{amssymb}
\usepackage{pifont}

\usepackage[colorinlistoftodos]{todonotes}

\usepackage{wrapfig}

\usepackage{cite}

\usepackage{amsthm}

\usepackage{titlesec}
\usepackage{titletoc}
\titleclass{\subsubsubsection}{straight}[\subsection]
\newcounter{subsubsubsection}[subsubsection]
\renewcommand\thesubsubsubsection{\thesubsubsection.\arabic{subsubsubsection}}
\titleformat{\subsubsubsection}{\normalfont\normalsize\bfseries}{\thesubsubsubsection}{1em}{}
\titlespacing*{\subsubsubsection}{0pt}{3.25ex plus 1ex minus .2ex}{1.5ex plus .2ex}
\titleformat*{\section}{\fontsize{11}{12}\selectfont\bfseries}
\titleformat*{\subsection}{\fontsize{11}{12}\selectfont\bfseries}
\titleformat*{\subsubsection}{\fontsize{11}{12}\selectfont\bfseries}
\titleformat*{\paragraph}{\fontsize{11}{12}\selectfont\bfseries}
\titleformat*{\subparagraph}{\fontsize{11}{12}\selectfont\bfseries}

\titlecontents{subsubsubsection}
[8em] 
{\small}
{\hyperlink{toc.\thecontentslabel}{\thecontentslabel.} }
{}
{\ \titlerule*[.5pc]{.}\contentspage}

\titlespacing*{\subsubsubsection}{0pt}{3.25ex plus 1ex minus .2ex}{1.5ex plus .2ex}

\usepackage{caption}
\usepackage{microtype}

\usepackage{enumitem}
\setlist{itemsep=0em}

\usepackage{fancyhdr} 
\usepackage{lastpage} 

\usepackage{listings}
\usepackage{xcolor}

\definecolor{codegreen}{rgb}{0,0.6,0}
\definecolor{codegray}{rgb}{0.5,0.5,0.5}
\definecolor{codepurple}{rgb}{0.58,0,0.82}
\definecolor{backcolour}{rgb}{0.95,0.95,0.92}

\lstdefinestyle{mystyle}{
    backgroundcolor=\color{backcolour},   
    commentstyle=\color{codegreen},
    keywordstyle=\color{magenta},
    numberstyle=\tiny\color{codegray},
    stringstyle=\color{codepurple},
    basicstyle=\ttfamily\footnotesize,
    breakatwhitespace=false,         
    breaklines=true,                 
    captionpos=b,                    
    keepspaces=true,                 
    numbers=left,                    
    numbersep=5pt,                  
    showspaces=false,                
    showstringspaces=false,
    showtabs=false,                  
    tabsize=2
}

\title{Time Series Transformer-Based Modeling of Pavement Skid and Texture Deterioration}

\author[1]{Lu Gao, Ph.D.}
\author[1]{Zia Ud Din, Ph.D.}
\author[1]{Kinam Kim, Ph.D.}
\author[1]{Ahmed Senouci, Ph.D.}

\affil[1]{Department of Civil and Environmental Engineering, University of Houston}
\date{}

\begin{document}

\maketitle    

\section*{Abstract}

This study investigates the deterioration of skid resistance and surface macrotexture following preventive maintenance using micro-milling techniques. Field data were collected from 31 asphalt pavement sections located across four climatic zones in Texas. The data encompasses a variety of surface types, milling depths, operational speeds, and drum configurations. A standardized data collection protocol was followed, with measurements taken before milling, immediately after treatment, and at 3, 6, 12, and 18 months post-treatment. Skid number and Mean Profile Depth (MPD) were used to evaluate surface friction and texture characteristics. The dataset was reformatted into a time-series structure with 930 observations, including contextual variables such as climatic zone, treatment parameters, and baseline surface condition. A comparative modeling framework was applied to predict the deterioration trends of both skid resistance and macrotexture over time. Eight regression models, including linear, tree-based, and ensemble methods, were evaluated alongside a time series transformer model. Results show that the transformer model achieved the highest prediction accuracy for skid resistance ($R^2=0.981$), while Random Forest performing best for macrotexture prediction ($R^2 = 0.838$). The findings indicate that the degradation of surface characteristics after preventive maintenance is nonlinear and influenced by a combination of environmental and operational factors. This study demonstrates the effectiveness of data-driven modeling in supporting transportation agencies with pavement performance forecasting and maintenance planning.\\

\noindent \textbf{Keywords}: Pavement deterioration, skid resistance, pavement performance modeling, transformer, time series, macrotexture, micro-milling, machine learning, deep learning, preventive maintenance

\section{Introduction}
Pavement surface characteristics such as skid resistance and macrotexture are crucial indicators for ensuring road safety and preventing accidents \citep{he2024skid}. Skid resistance, the friction generated between tires and the pavement surface, is fundamental for vehicle control during critical maneuvers like turning and emergency braking \citep{canestrari2024use}. A significant decrease in skid resistance can lead to a considerable increase in accident rates, and higher skid numbers are consistently associated with fewer crashes \citep{subedi2025safety}. Macrotexture, defined as the surface texture component with wavelengths between 0.5 and 50 mm, plays a vital role in removing water film from the tire-pavement contact area and contributes to skid resistance, particularly at higher speeds and in wet conditions \citep{hibbs1996tire}. Both microtexture (roughness less than 0.5 mm) and macrotexture together determine the overall pavement friction. Pavement skid resistance undergoes a natural progression of change over time due to traffic and environmental influences. It may initially increase as the asphalt binder wears away, then decline as the aggregate becomes polished, and eventually reach a stable state \citep{kumar2021review}. Therefore, consistent monitoring and maintenance of these pavement surface characteristics are essential to maintain adequate traction and mitigate crash risks, especially under wet weather conditions \citep{yang2024field}.

Preventive maintenance techniques, such as micro-surfacing and milling operations, are applied to significantly alter existing pavement surface properties \citep{dave2025understanding}. For instance, micro-surfacing aims to improve anti-skid performance, wear resistance, and durability. It also addresses noise reduction through optimized gradation and the inclusion of additives like rubber powder and fiber \citep{li2024design}. However, the effectiveness of such treatments remains insufficiently quantified. This study addresses the following research question: How do skid resistance and surface macrotexture deteriorate over time after micro-milling treatments; and to what extent can sequence-based machine learning models improve the prediction of this degradation compared to traditional approaches?

\section{Literature Review}

\subsection{The impact of skid resistance and surface texture on road safety}
Pavement skid resistance is a critical factor for ensuring road safety and preventing accidents \citep{khasawneh48investigating}. Insufficient skid resistance is identified as a significant cause of road traffic crashes, accounting for approximately 30\% of such incidents worldwide annually \citep{he2024skid}. While newly constructed pavements generally exhibit high skid resistance, this property continuously deteriorates over time due to the repeated actions of traffic loads and environmental factors \citep{li2024review}. For asphalt pavements, the deterioration typically follows a pattern: an initial increase in friction as the asphalt binder is stripped, followed by a decline as the aggregate surface polishes, and finally stabilizes in the late period \citep{do2009physical}.

The ability of a pavement to resist skidding is heavily influenced by its surface texture, which is categorized into micro-texture and macro-texture \citep{canestrari2024use}. Micro-texture refers to the fine-scale roughness of the aggregate surface, with wavelengths less than 0.5 mm, and primarily contributes to adhesion and friction at low speeds \citep{de2019locked}. Macro-texture, on the other hand, involves larger irregularities with wavelengths between 0.5 mm and 50 mm, which are crucial for hysteretic phenomena, effective water drainage, and maintaining skid resistance at higher speeds \citep{tian2020pavement}. The presence of water or other contaminants on the pavement surface complicates the role of texture by reducing the tire-pavement contact area and building up hydrodynamic pressure, thereby increasing the risk of skidding and hydroplaning \citep{fwa2024pavement}.

A strong correlation exists between pavement surface friction values and road safety. Higher skid resistance usually linked to reduced crash frequencies, particularly under wet conditions \citep{subedi2025safety}. Studies have shown that a 10-unit increase in Skid Number (SN) can lead to a 7\% to 8\% reduction in dry-weather crashes and a more significant 14\% to 21\% decrease in wet-weather crashes on highways \citep{mayora2009assessment}. The impact of increased skid resistance is especially pronounced on interstate highways and curved segments during wet conditions, where friction demand is higher \citep{long2014quantitative}. 

\subsection{Influence of micro-milling treatment on skid resistance and surface texture}
Micro-milling is a cost-effective pavement preservation technique widely adopted in highway maintenance and rehabilitation \citep{tsai2019evaluation}. This method involves removing a thin layer of existing asphalt pavement, typically as little as 3/8 inch (9.5 mm) or 1.5–2 cm for tunnel concrete pavement, using a cutting drum equipped with numerous small, tightly-spaced teeth \citep{gao2015milled}. The primary goals of micro-milling include improving skid resistance, restoring the road's profile, and removing surface distresses such as raveling or minor rutting before applying a new overlay. The micro-milled surface can be opened directly to traffic, giving contractors greater flexibility \citep{nittinger1977milling}.

The effectiveness of micro-milling on surface texture and skid resistance is significantly influenced by operational factors. Sections milled with fine drums, typical of micro-milling, demonstrate enhanced skid resistance and macrotexture \citep{dave2025understanding}. Higher forward milling speeds also correlate positively with improved skid resistance and macrotexture. The recommended speeds ranges from 70 to 80 feet per minute \citep{gao2015milled}. Optimal cutting depths for texturing are typically between 0.25 and 0.5 inches (6 to 13 mm). However, the milling depth itself does not notably impact the long-term deterioration of skid resistance or macrotexture \citep{cho2024performance}. 


Micro-milling faces challenges in achieving consistent texture and depth. Factors such as existing pavement distress, machine performance, and drum rotation rates can influence these outcomes \citep{hui2022quality}. Inconsistent milling can compromise bonding strength, leading to premature distress such as peeling or shoving in the overlay \citep{hung2014effects}. Micro-milling has demonstrated service lives ranging from about 12 months on seal coats to 10-12 years on hot-mix asphalt (HMA) sections \citep{khanal2023ultra}.

\subsection{Factors affecting skid resistance and surface texture deterioration}

Pavement skid resistance is not static. It continuously deteriorates over time due to various influencing factors, primarily traffic loads and environmental conditions \citep{wei2022study}. The deterioration process usually occurs in stages: first, a rapid increase due to binder removal, then a slow decrease as aggregate micro-texture worsens, and finally, a stabilization phase \citep{unal2021effect}. Vehicular traffic, through its polishing effect, is a major contributor to this decline, especially in the wheel path areas \citep{kane2022new}. Vehicle movement determines the direction of polishing. Acceleration-dominated areas face higher friction against traffic flow, while deceleration-dominated areas experience lower friction in the direction of traffic flow \citep{chu2022directional}. Environmental factors, including water, snow, ice, and contaminants like rubber deposits, significantly reduce skid resistance. They do this by forming a lubricating film or covering the surface texture \citep{persson2005rubber}. Water film thickness, in particular, negatively correlates with skid resistance and can lead to hydroplaning at higher speeds. Temperature (both tire and ambient/pavement) can also influence skid resistance, with increases often leading to a reduction \citep{baran2011temperature}.

Beyond traffic and environmental effects, specific material and design choices, such as the use of composite aggregates, significantly impact skid resistance and its long-term performance \citep{peng2024long, buildings14082339}. The mineralogy of aggregates greatly influences their susceptibility to wear and polishing. For example, poly-mineral aggregates tend to exhibit higher polishing resistance and friction than mono-mineral ones like limestone \citep{zhang2024impact}. Aggregate dimension also plays a role, with finer aggregates typically providing higher initial friction but experiencing a more pronounced reduction with polishing \citep{roshan2025impact}. For asphalt mixtures, the long-term skid resistance is primarily determined by coarse aggregates, while fine aggregates mainly affect the initial friction \citep{liu2021effect}. 

The type and dosage of binder can influence the initial stripping phase, which affects both friction values and the time required for complete binder removal \citep{alkofahi2019evaluation}. Pavement macro-texture and air voids are crucial, as they influence the actual tire-pavement contact area and drainage capacity \citep{pranjic2018pavement}. The use of recycled materials like reclaimed asphalt pavement (RAP) or crumb rubber, and warm mix asphalt (WMA) additives, may have a detrimental effect on skid resistance and can delay binder removal \citep{putra2019skid}. Grooving pavements improves drainage and skid resistance, and reduces hydroplaning risk. However, groove deterioration from wear or rubber deposits can lessen their effectiveness \citep{pasindu2020analytical}. Pavement markings also influence skid resistance. The materials used and glass beads can decrease friction compared to the surrounding pavement. However, anti-skid additives can help to reduce this effect. \citep{anderson1980wet}.

\subsection{Deterioration modeling of skid Resistance and surface texture}

Modeling pavement skid resistance and texture deterioration is vital for road safety and maintenance. It helps understand the complex interactions between vehicle tires and road surfaces \citep{chen2025review}. Various analytical, statistical, and machine learning models have been developed to predict and understand these time-dependent properties \citep{kumar2023state,gao2012bayesian,archilla2000development}.

Statistical and analytical models have traditionally been used to describe the decay of pavement conditions \citep{gao2011performance,zhang2018nested,gao2017bayesian}. Common approaches include empirical relationships and regression models. These link pavement characteristics to skid resistance or its decline rate. Characteristics used include texture parameters like Mean Texture Depth (MTD) and Mean Profile Depth (MPD) \citep{hu2022evaluate}. Models like the Penn State model, PIARC International Friction Index (IFI), and the Rose and Gallaway model are examples that estimate skid resistance as a function of speed and texture, though they often have limitations regarding water film thickness or application scope \citep{fwa2021determination}. Skid resistance often declines rapidly at first, then slows, eventually stabilizing. Models like asymptotic, exponential, and logarithmic ones capture this behavior \citep{du2021promoting}. However, these statistical methods may struggle with non-linear relationships and the multitude of influencing factors, leading to limited accuracy and applicability \citep{ji2022establishment}.

Machine learning (ML) has emerged as a powerful tool for predicting pavement conditions. It overcomes the limitations of traditional statistical models by handling complex, non-linear relationships and multi-dimensional data \citep{kone2023application,peng2025evaluating,gao2024considering,yu2023pavement,gao2021detection}. Widely applied ML algorithms include Artificial Neural Networks (ANN), Support Vector Machines (SVM), Random Forest (RF), and Decision Trees (DT) \citep{kone2023application}. These models leverage various input parameters, such as vehicle speed, water depth, tire tread depth, pavement macro-texture, aggregate properties, and even environmental factors, to predict friction coefficients or classify skid resistance levels (e.g., low, medium, high) \citep{rajabipour2021use}. Data from 3D laser scanning and other non-contact measurement methods provide high-resolution texture information as input for these ML models, improving prediction accuracy \citep{li2024review}.

More advanced AI and ensemble learning models are continually being developed to enhance prediction accuracy and interpretability \citep{weng2025interpretable,gao2024considering}. Neural networks optimized by genetic algorithms demonstrate strong self-learning capabilities. They can establish highly non-linear relationships for skid resistance prediction. This applies even to long-term analysis and specific pavement types, such as steel bridge deck pavement \citep{liu2022developing}. Light Gradient Boosting Machine (LightGBM) and its optimized variant, Bayesian-LightGBM, are known for their stability, generalization, and efficiency in multi-factor regression. They significantly improve prediction accuracy over other models \citep{hu2022evaluate}. 

Deep learning approaches, such as Convolutional Neural Networks (CNN) and deep residual networks (ResNets), are also being explored, particularly with 3D texture data, to achieve higher accuracy in correlating texture with friction \citep{li2024review,gao2023deep}. Integrated frameworks that combine techniques such as Fast Fourier Transform (FFT) with Extreme Gradient Boosting (XGBoost) or wavelet transforms with fractal theory enhance texture characterization and improve skid resistance prediction. These frameworks provide insights into the impact of multi-scale texture features \citep{zhan2022integrated}.

\subsection{Limitations and Research Gaps}

While several studies have investigated the degradation of pavement skid resistance and surface texture, important limitations remain, especially in the context of their evolution after preventive maintenance treatments. Most existing models rely on traditional statistical or machine learning approaches that treat each data point as independent. As a result, they fail to consider the temporal continuity and dynamic progression of surface performance over time. These approaches often overlook how previous surface conditions and operational factors influence future outcomes, leading to reduced predictive accuracy in long-term assessments. There is a clear need for models that can capture time-dependent patterns and complex relationships among treatment parameters, environmental conditions, and surface characteristics. This study addresses this gap by proposing a sequence-based Transformer model that treats the degradation of skid resistance and texture after maintenance as a temporal learning problem.

\section{Methodology}

\subsection{Problem Formulation}

Predicting the temporal evolution of pavement skid resistance is a classic sequence–to–one time series regression task. In this task, the model uses a sequence of past pavement conditions to predict what the road will be like at the next inspection. For example, if the past five years of data are available, the model learns patterns from those five years and uses them to forecast the skid resistance and texture for the sixth year. For a given test section, a finite sequence of feature vectors  $\mathbf{x}_{0},\mathbf{x}_{1},\dots,\mathbf{x}_{T}$ is observed at discrete time points $t_0, t_1, \dots, t_T$, where each $\mathbf{x}_{t} \in \mathbb{R}^{d_x}$ represents the concatenation of all feature values observed at time $t$. Given a sliding window of length $L$, the Transformer receives the ordered sequence
\begin{equation}
  \mathbf{X}_{t-L+1:t}\;=\;
  \begin{bmatrix}
     \mathbf{x}_{t-L+1}\\ \vdots \\ \mathbf{x}_{t}
  \end{bmatrix} \in \mathbb{R}^{L\times d_x}
\end{equation}
as input and outputs a forecast $\hat{y}_{t+\Delta t} \in \mathbb{R}$ of the skid number at the next inspection point $t+\Delta t$.

To capture both short- and long-term temporal dependencies in this sequence-to-one prediction task, the Transformer model \citep{vaswani2017attention} was adopted.  Its self-attention mechanism makes it well-suited for time series forecasting, especially when handling long-range dependencies and multivariate inputs.

\subsection{Time Series Transformer}

The Transformer architecture, originally developed for natural language processing, is well suited for time series forecasting tasks with complex temporal dependencies. Unlike traditional models such as ARIMA or recurrent neural networks (RNNs), which process sequences sequentially, the transformer utilizes a self-attention mechanism that considers all time steps simultaneously. This enables the model to more effectively capture long-range dependencies and interactions among variables across time \citep{wen2022transformers}. 

Time Series transformers adapt the original transformer architecture to forecasting problems by replacing word embeddings with time series feature vectors and modifying the positional encoding to reflect temporal structure. These models are capable of handling multivariate inputs, irregular time steps, and long input sequences. Time series transformers have been widely applied in multivariate forecasting tasks \citep{zerveas2021transformer,li2019enhancing,jin2022time}. The following section introduces the key concepts of the Transformer architecture.

\subsubsection{Feature Embedding and Positional Encoding}

The Transformer is a neural architecture designed for sequence modeling. The Transformer architecture processes all inputs within a common latent space of fixed dimensionality $d_{\text{model}}$. This design allows the model to uniformly apply self-attention and feed-forward layers regardless of the input's original structure or scale. To accommodate this, each input vector $\mathbf{x}_{t} \in \mathbb{R}^{d_x}$ is first projected into the latent space through a learned linear transformation \citep{vaswani2017attention}:
\begin{equation}
  \mathbf{h}_{t}^{(0)} \;=\; \mathbf{E}\,\mathbf{x}_{t}, \quad \mathbf{E} \in \mathbb{R}^{d_{\text{model}} \times d_x}.
\end{equation}
The resulting vector $\mathbf{h}^{(0)}_{t}$ serves as the initial token representation for time $t$, prior to any self-attention processing.

To retain the temporal order of the sequence, a positional encoding vector $\mathbf{p}_t \in \mathbb{R}^{d_{\text{model}}}$ is added to each embedding. This positional signal distinguishes tokens from different time steps, even if their feature content is similar. A widely used deterministic choice is the sinusoidal encoding \citep{vaswani2017attention}:

\begin{equation}  
  \bigl(\mathbf{p}_{t}\bigr)_{2k}=\sin\!\Bigl(\tfrac{t}{10000^{2k/d_{\text{model}}}}\Bigr),
  \qquad
  \bigl(\mathbf{p}_{t}\bigr)_{2k+1}=\cos\!\Bigl(\tfrac{t}{10000^{2k/d_{\text{model}}}}\Bigr),
\end{equation}
for $k = 0,\dots,\lfloor d_{\text{model}}/2\rfloor - 1$. This scheme encodes different frequencies along dimensions of the embedding, allowing the model to represent both coarse and fine temporal patterns. The final input to the encoder is a sequence of order-aware embeddings formed by addition:
\begin{equation}    
  \tilde{\mathbf{h}}_{t}^{(0)} = \mathbf{h}_{t}^{(0)}+\mathbf{p}_{t},
  \quad\;
  \mathbf{H}^{(0)}
  =
  \bigl[\tilde{\mathbf{h}}_{t-L+1}^{(0)},\dots,\tilde{\mathbf{h}}_{t}^{(0)}\bigr] \in \mathbb{R}^{L \times d_{\text{model}}}
\end{equation}
The addition (rather than concatenation) ensures that the input dimensionality to the encoder remains fixed at $d_{\text{model}}$.

\subsubsection{Self–Attention Encoder}

Each Transformer encoder layer consists of two main components: multi-head self-attention, which allows the model to relate each time step to every other time step in the input sequence, and a position-wise feed-forward network (FFN), which processes each time step independently through a small neural network to enhance the model’s capacity. To make training more stable and efficient, the Transformer adds the original input back to the output of each part (a technique called a residual connection) and then applies a method called layer normalization, which adjusts the values to be more balanced and consistent. These techniques help the model learn better and avoid problems like exploding or vanishing gradients.

Let $\mathbf{H}^{(l)} \in \mathbb{R}^{L \times d_{\text{model}}}$ be the input to layer $l$. Following the scaled‑dot attention of \citet{vaswani2017attention}, the self-attention mechanism computes query, key, and value matrices:
\begin{equation}
\mathbf{Q} = \mathbf{H}^{(l)} \mathbf{W}_Q,\quad
\mathbf{K} = \mathbf{H}^{(l)} \mathbf{W}_K,\quad
\mathbf{V} = \mathbf{H}^{(l)} \mathbf{W}_V,
\end{equation}
where $\mathbf{W}_Q, \mathbf{W}_K, \mathbf{W}_V \in \mathbb{R}^{d_{\text{model}} \times d_k}$ are learned parameters. Attention weights are computed via scaled dot-product:
\begin{equation}
\mathbf{A} = \operatorname{softmax}\left(\frac{\mathbf{Q} \mathbf{K}^\top}{\sqrt{d_k}}\right) \in \mathbb{R}^{L \times L},
\end{equation}
and the attention output is
\begin{equation}
\operatorname{Attn}(\mathbf{Q}, \mathbf{K}, \mathbf{V}) = \mathbf{A} \mathbf{V} \in \mathbb{R}^{L \times d_k}.
\end{equation}

Using $H$ independent attention heads allows the model to capture diverse interaction patterns. Their outputs are concatenated and linearly projected back \citep{vaswani2017attention}:
\begin{equation}
\operatorname{MHA}^{(l)}(\mathbf{H}^{(l)}) = 
\left[\operatorname{Attn}^{(1)} \Vert \dots \Vert \operatorname{Attn}^{(H)}\right] \mathbf{W}_O,
\end{equation}
with $\mathbf{W}_O \in \mathbb{R}^{H d_k \times d_{\text{model}}}$.

Each sub-layer includes a residual connection and layer normalization:
\begin{equation}
\tilde{\mathbf{H}}^{(l)} = \operatorname{LayerNorm}\left(\mathbf{H}^{(l)} + \operatorname{MHA}^{(l)}(\mathbf{H}^{(l)})\right).
\end{equation}

Following \citet{vaswani2017attention}, a feed-forward network is then applied to each time step independently:
\begin{equation}
\operatorname{FFN}(\mathbf{h}) = \sigma(\mathbf{h} \mathbf{W}_1 + \mathbf{b}_1) \mathbf{W}_2 + \mathbf{b}_2,
\end{equation}

Using residual connections \citep{he2016deep} and layer normalization \citep{ba2016layer}, the final output of the encoder layer is:
\begin{equation}
\mathbf{H}^{(l+1)} = \operatorname{LayerNorm}\left(\tilde{\mathbf{H}}^{(l)} + \operatorname{FFN}(\tilde{\mathbf{H}}^{(l)})\right).
\end{equation}

Stacking $N$ such layers yields the final output $\mathbf{H}^{(N)}$, which captures contextual information from the entire input sequence.

\subsubsection{Sequence Pooling and Regression Head}

Following common practice in Transformer-based regression models (e.g., \citet{vaswani2017attention}),  the final time-step embedding $\mathbf{h}_{t}^{(N)}$ is extracted and passed through a small multi-layer perceptron.
\begin{equation}
  \hat{y}_{t+\Delta t} \;=\; 
  \mathbf{w}_{2}^{\!\top}\sigma\!\bigl(\mathbf{W}_{1}\mathbf{h}_{t}^{(N)}+\mathbf{b}_{1}\bigr)+b_{2}.
\end{equation}

\subsection{Loss Function and Optimisation}

The network parameters $\Theta$ are trained to minimise the mean-squared error, commonly used in regression tasks \citep{goodfellow2016deep}. 
\begin{equation}
  \mathcal{L}(\Theta)\;=\;
  \frac{1}{N}\sum_{i=1}^{N}\bigl(\,y_{i}-\hat{y}_{i}\bigr)^{2},
\end{equation}
where $N$ is the number of (sequence, target) pairs in the training set.  
Stochastic gradient descent with adaptive learning rate is typically used, coupled with early stopping on a validation set.

\section{Case Study}

\subsection{Data Collection}

\begin{figure}[H]
    \centering
    \captionsetup{justification=centering}
    \includegraphics[width=0.6\linewidth]{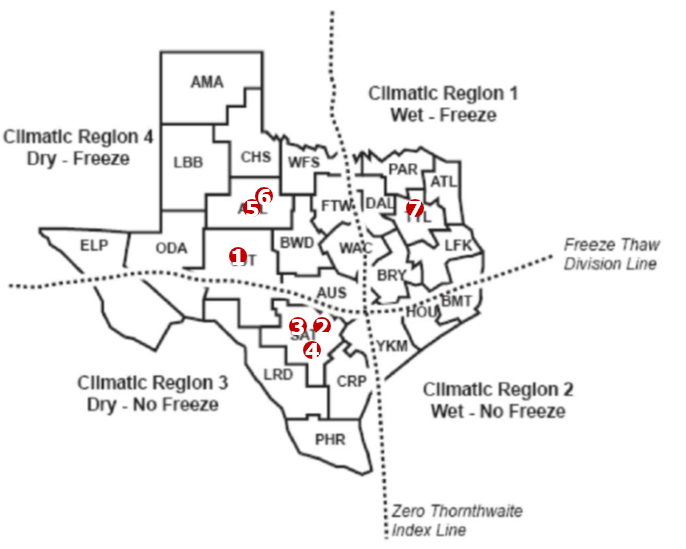}
    \caption{Test Sits}
    \label{fig:sites}

    \vspace{0.5em}
    {\small \textit{Note:} 1. US 87 (SJT); 2. Wurzbach Parkway (SAT); 3. SH 151-1 (SAT); 4. SH 151-2 (SAT); 5. US 277-1 (ABL); 6. US 277-2 (ABL); 7. I-20 (TYL)
}
    
\end{figure}

The data used in this study were obtained from a research initiative conducted by the Texas Department of Transportation (TxDOT) to evaluate the short-term effectiveness of micro-milling in improving pavement skid resistance. The study involved 31 pavement sections across multiple regions of Texas, with repeated measurements collected over time to monitor changes in surface conditions. The selected test sections span multiple climatic zones across the state of Texas (Figure \ref{fig:sites}). These sections are distributed across various TxDOT districts, including San Angelo, San Antonio, Abilene, and Tyler. In total, the study includes 16 seal coat sections and 15 HMA sections located on major roadways such as US 87, SH 151, US 277, I-20 West, and Wurzbach Parkway.

\begin{figure}[H]
\centering

\subfloat[US87]{%
  \includegraphics[width=0.3\linewidth]{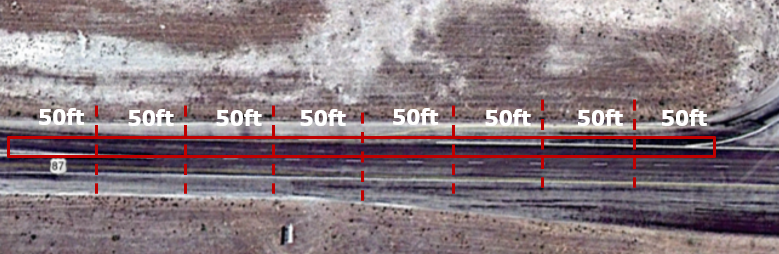}
  \label{fig:us87_layout}
}
\hfill
\subfloat[Wurzbach Parkway]{%
  \includegraphics[width=0.3\linewidth]{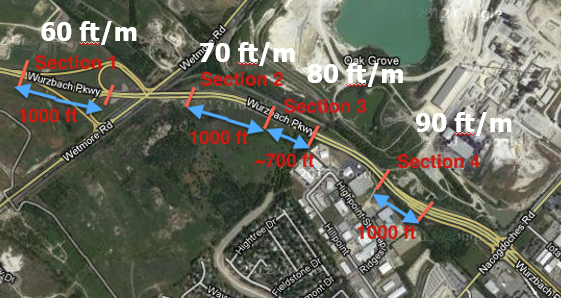}
  \label{fig:wp_layout}
}
\hfill
\subfloat[SH151]{%
  \includegraphics[width=0.3\linewidth]{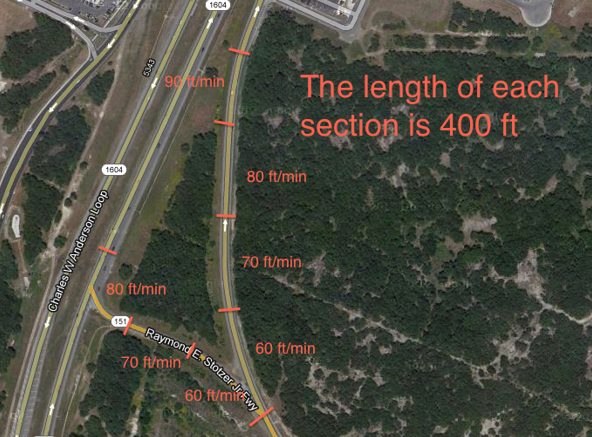}
  \label{fig:sh151_layout}
}

\vspace{1em}

\subfloat[US277 (Phase I)]{%
  \includegraphics[width=0.45\linewidth]{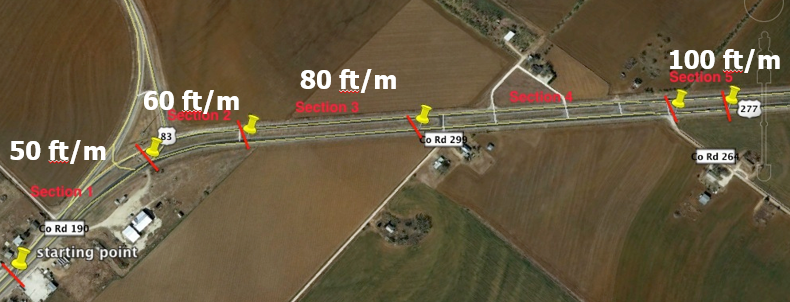}
  \label{fig:us277_layout}
}
\hfill
\subfloat[US277 (Phase II)]{%
  \includegraphics[width=0.45\linewidth]{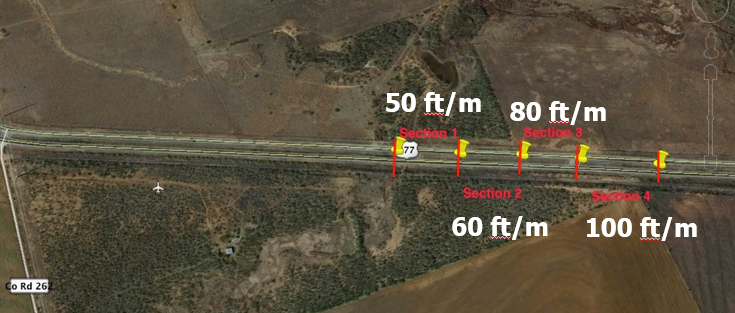}
  \label{fig:us277_2_layout}
}

\caption{Experiment section layouts at five locations}
\label{fig:layout_all}
\end{figure}

Figures~\ref{fig:us87_layout} to~\ref{fig:us277_2_layout} illustrate the layouts of selected experiment sections across several highway corridors in Texas. These layouts represent a subset of the full set of test segments used in this study, chosen to highlight typical field configurations and treatment variations. Figure~\ref{fig:us87_layout} shows the configuration of a test site along US87, where uniformly spaced short segments were prepared to investigate short-term changes under consistent baseline conditions. This site enabled detailed tracking of texture and friction immediately after milling. The Wurzbach Parkway layout, shown in Figure~\ref{fig:wp_layout}, includes longer segments with varying milling speeds and depths, designed to capture the influence of operational parameters on surface evolution. Figure~\ref{fig:sh151_layout} illustrates another configuration on SH151. Two sets of experimental sections along US277 are depicted in Figures~\ref{fig:us277_layout} and~\ref{fig:us277_2_layout}. These segments span a broader range of milling depths and were constructed during two separate phases. 

Beyond segment configurations, Figure \ref{fig:layout_all} also emphasizes the geographic and contextual diversity of the experiment sites. The five locations span urban and rural corridors, including major highways and regional roads across different TxDOT districts. This geographic spread allows the study to capture a broader range of traffic loading conditions, construction practices, and environmental exposures—factors that significantly influence surface deterioration. By including sites such as Wurzbach Parkway in a metropolitan area and US277 in rural regions, the dataset used in this case study supports more generalizable modeling and enables performance comparisons across varying climatic zones, pavement types, and maintenance practices. 






\begin{figure}[H]
\centering

\subfloat[Roadway before milling operation\label{fig:before_after_a}]{
    \includegraphics[width=0.38\linewidth]{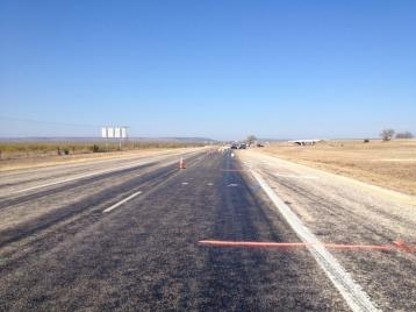}
}
\hfill
\subfloat[Milling machinery in operation\label{fig:before_after_b}]{
    \includegraphics[width=0.38\linewidth]{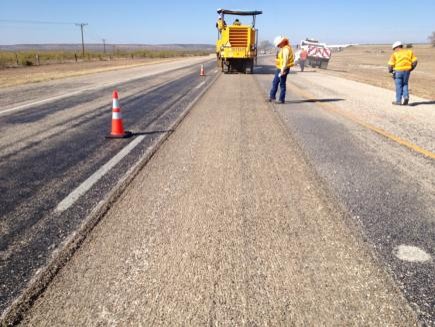}
}

\caption{Milling Operation on Roadway}
\label{fig:before_after}
\end{figure}

To improve surface texture and skid resistance, surface treatments were applied using either standard milling drums with approximately 150 teeth or fine-milling drums with around 300 teeth. Milling depths ranged from 0.2 to 0.5 inches, and machine operating speeds varied between 30 and 100 feet per minute. Each pavement section underwent a controlled surface texturing treatment, during which engineers recorded key operational parameters such as milling depth, drum type, and machine speed. Representative field photos of the roadway before treatment and the milling process in operation are shown in Figure~\ref{fig:before_after}, which illustrates key stages of the micro-milling process. Figure \ref{fig:before_after_a} shows the pavement surface prior to treatment. Visually, the surface appears uniformly dark and smooth, with a noticeable sheen caused by aggregate polishing over time. This shiny black coloration is a common indicator of severely reduced macrotexture and microtexture, which suggests low skid resistance. Such conditions can significantly impair tire-pavement friction, especially in wet weather, and increase crash risks. Figure~\ref{fig:before_after_b} illustrates the micro-milling process, during which the worn and polished surface layer is removed to restore surface texture. This operation exposes a rougher, lighter-colored aggregate matrix, which improves macrotexture and enhances skid resistance.






\begin{figure}[H]
\centering

\begin{minipage}{0.3\linewidth}
    \centering
    \includegraphics[width=0.8\linewidth]{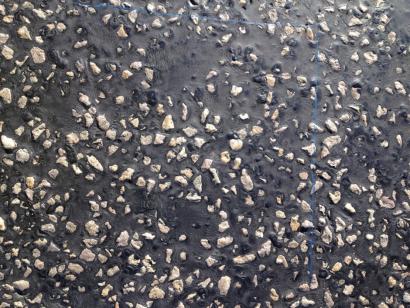}
    \caption*{(a) US87 before milling}
\end{minipage}
\hfill
\begin{minipage}{0.3\linewidth}
    \centering
    \includegraphics[width=0.8\linewidth]{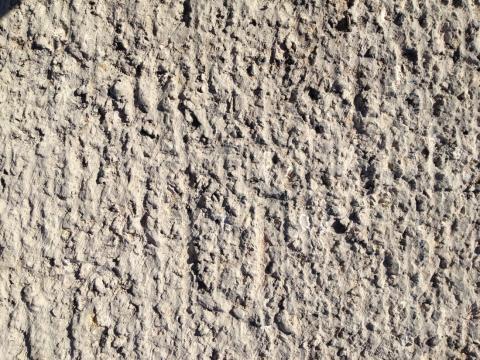}
    \caption*{(b) US87 immediately after milling}
\end{minipage}
\hfill
\begin{minipage}{0.3\linewidth}
    \centering
    \includegraphics[width=0.8\linewidth]{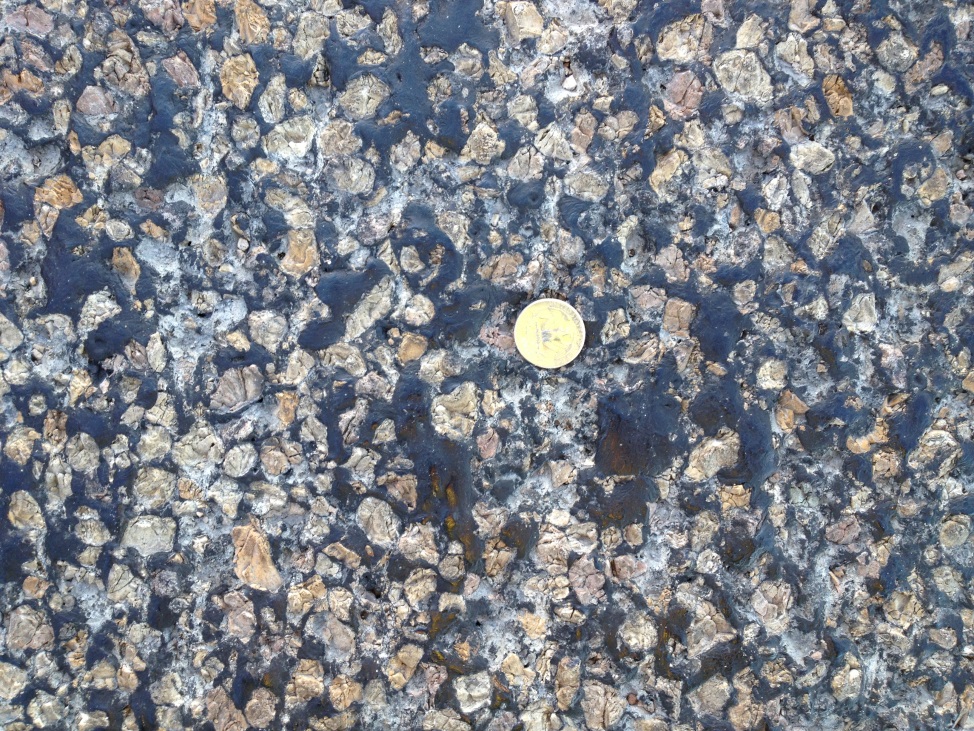}
    \caption*{(c) US87 3 months after milling}
\end{minipage}

\vspace{1em}

\begin{minipage}{0.3\linewidth}
    \centering
    \includegraphics[width=0.8\linewidth]{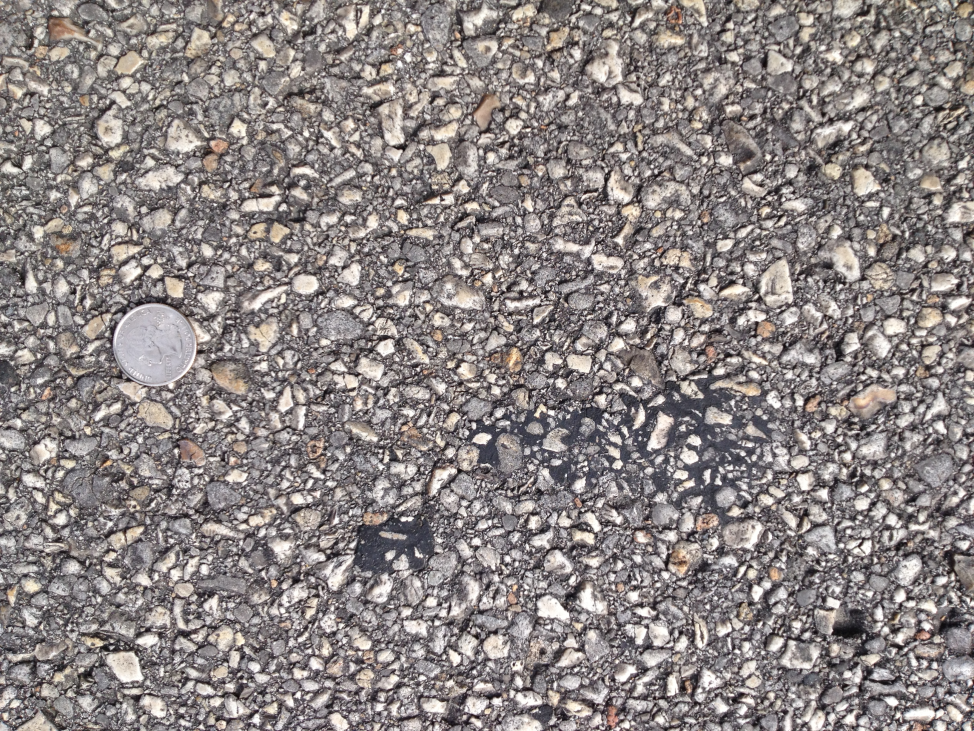}
    \caption*{(d) Wurzbach Pkwy before milling}
\end{minipage}
\hfill
\begin{minipage}{0.3\linewidth}
    \centering
    \includegraphics[width=0.8\linewidth]{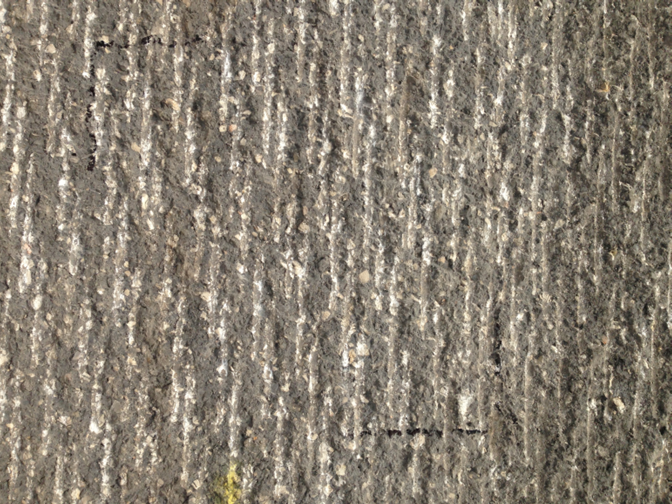}
    \caption*{(e) Wurzbach Pkwy immediately after milling}
\end{minipage}
\hfill
\begin{minipage}{0.3\linewidth}
    \centering
    \includegraphics[width=0.8\linewidth]{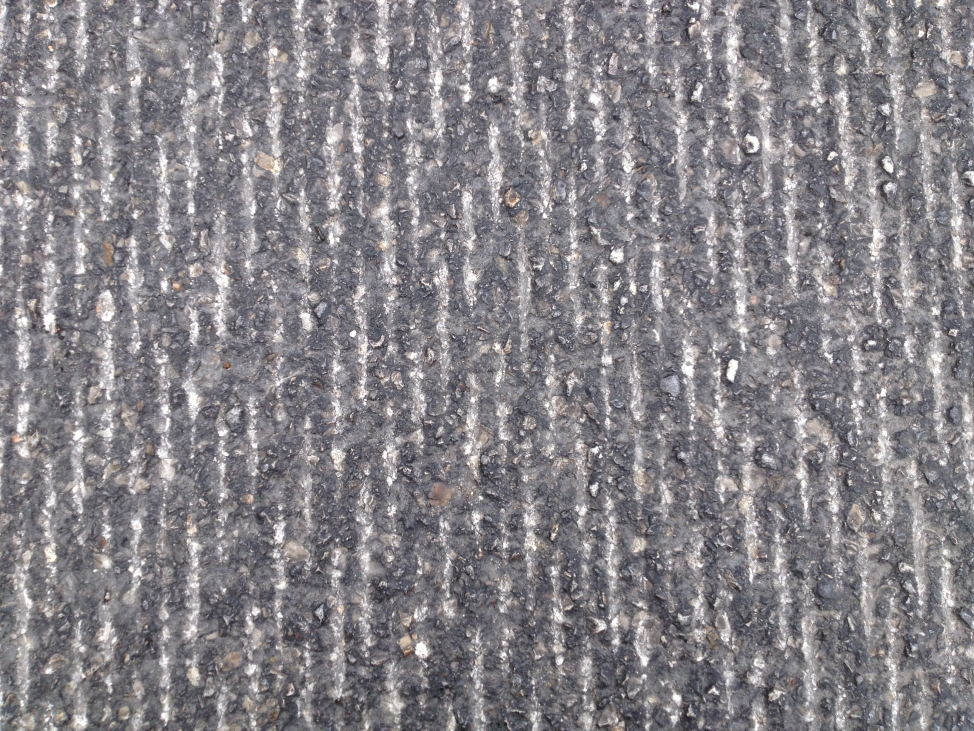}
    \caption*{(f) Wurzbach 3 months after milling}
\end{minipage}

\vspace{1em}

\begin{minipage}{0.3\linewidth}
    \centering
    \includegraphics[width=0.8\linewidth]{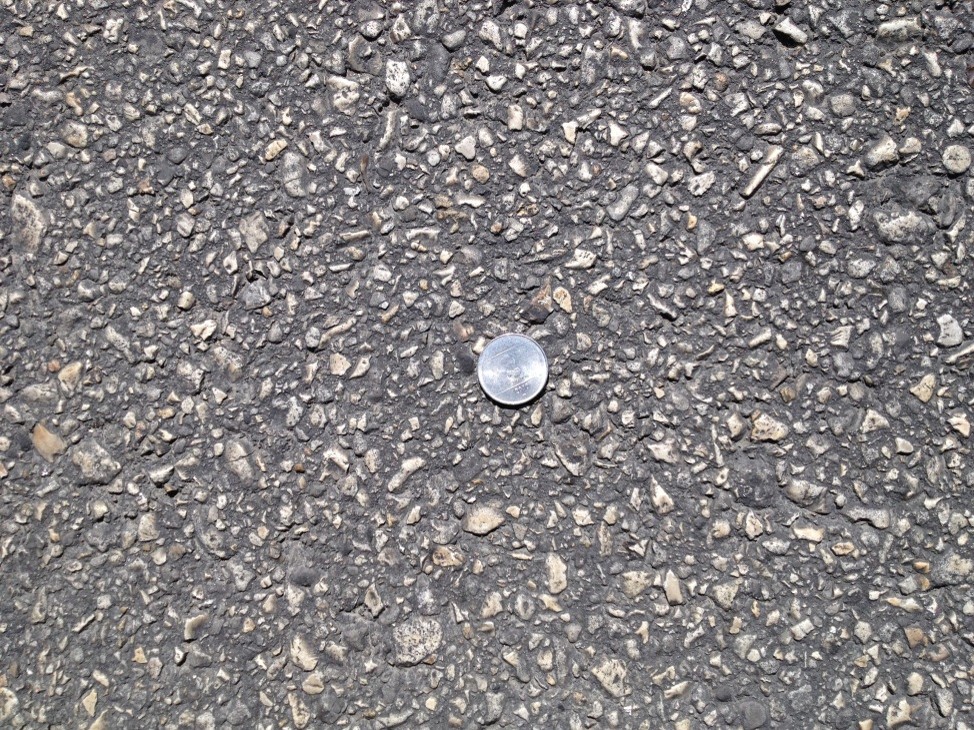}
    \caption*{(g) SH151 before milling}
\end{minipage}
\hfill
\begin{minipage}{0.3\linewidth}
    \centering
    \includegraphics[width=0.8\linewidth]{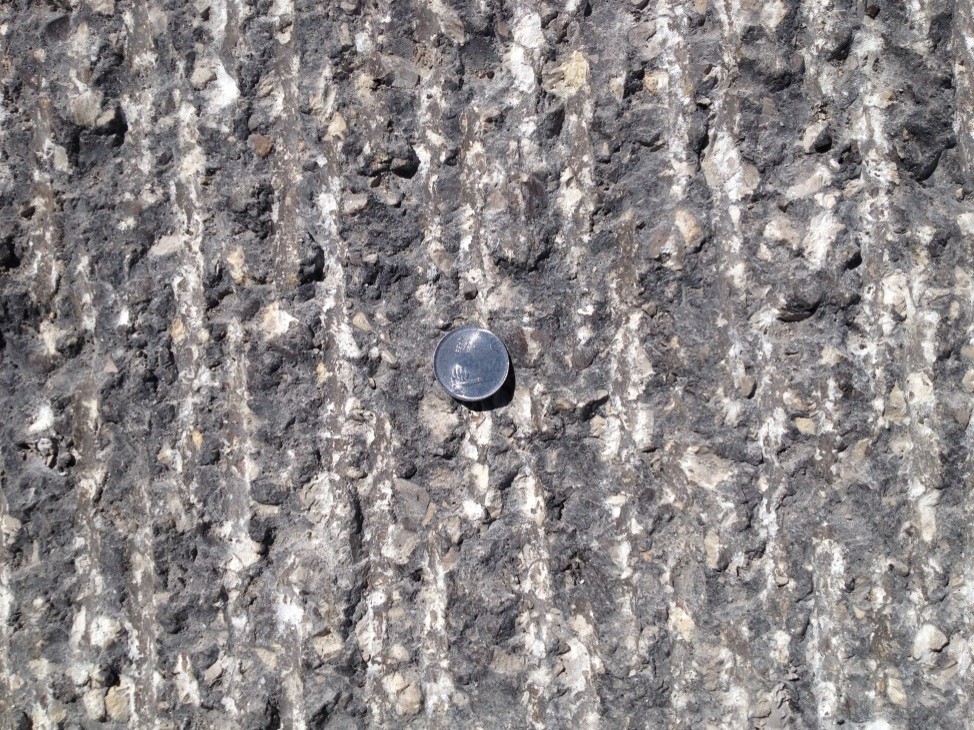}
    \caption*{(h) SH151 immediately after milling}
\end{minipage}
\hfill
\begin{minipage}{0.3\linewidth}
    \centering
    \includegraphics[width=0.8\linewidth]{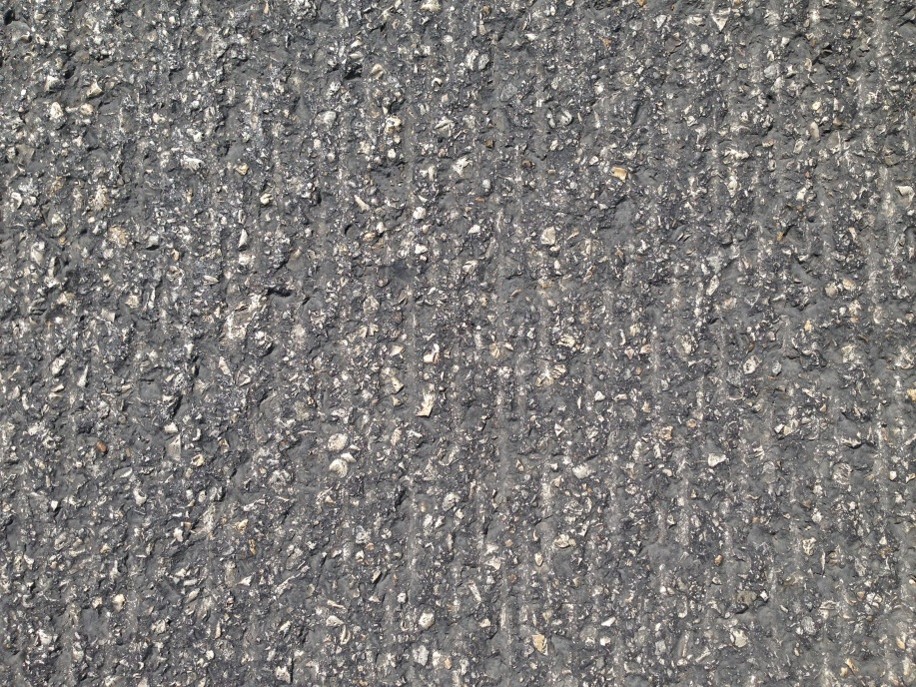}
    \caption*{(i) SH151 3 months after milling}
\end{minipage}

\caption{Pavement surfaces before, after, and 3 months after milling at three selected locations}
\label{fig:milling_comparison}
\end{figure}

Following treatment, each section was monitored at six scheduled time points: a baseline measurement before treatment, and follow-up measurements at approximately 0, 3, 6, 12, and 18 months after treatment. Figure \ref{fig:milling_comparison} displays the pavement surface before, immediately after, and three months following micro-milling at three chosen sites. At every time point, three replicate measurements were taken across the left, center, and right wheel paths of each lane to ensure spatial representativeness and data reliability.

The pre-treatment surfaces at US87 (Figure~\ref{fig:milling_comparison}a), Wurzbach Parkway (Figure~\ref{fig:milling_comparison}d), and SH151 (Figure~\ref{fig:milling_comparison}g) exhibit varying degrees of aggregate polishing and binder flushing. In particular, the US87 section demonstrates a noticeably darker and smoother surface, indicative of low friction due to binder bleeding and reduced macrotexture. The Wurzbach and SH151 sections show moderately exposed aggregates; however, the presence of surface fines suggests macrotexture infill and diminished surface drainage capability. 

Immediately after treatment, the images reveal a substantial improvement in surface texture uniformity and roughness. For instance, the US87 section (Figure~\ref{fig:milling_comparison}b) displays an open-textured and lighter-colored surface, which indicates successful removal of the oxidized top layer. The Wurzbach Parkway and SH151 sections (Figures~\ref{fig:milling_comparison}e and \ref{fig:milling_comparison}h) exhibit well-defined milling grooves and consistent texturing, reflective of a uniform cutting depth across the lane width. The increased surface roughness at this stage aligns with the immediate post-treatment rise in skid resistance observed in field data.

At three months post-treatment, changes in surface condition begin to emerge. The US87 section (Figure~\ref{fig:milling_comparison}c) shows evidence of re-polishing, with aggregate edges becoming less distinct, which suggests early-stage wear-in under traffic loading. In contrast, the textures at Wurzbach Parkway and SH151 (Figures~\ref{fig:milling_comparison}f and \ref{fig:milling_comparison}i) remain relatively well-preserved, with groove patterns still clearly visible. This difference may be attributed to variations in traffic volume, aggregate hardness, or binder type.

During each monitoring event, several variables were recorded. Skid resistance was assessed using both a locked-wheel skid trailer and a British Pendulum Tester. Surface texture was evaluated through the Sand Patch Test, Circular Texture Meter (CTM), and 3D laser scanning, yielding indicators such as Mean Profile Depth (MPD). Operational parameters, including milling depth, drum configuration, and machine speed, were also documented.

\subsection{Data Description}

The final dataset comprises 324 samples, each representing a specific lane segment at a particular time point during the monitoring period. These samples were collected from multiple pavement sections located in different climatic regions across Texas. The dataset incorporates a wide range of information, including skid resistance measurements, surface texture indicators, environmental conditions, and operational parameters from the surface treatment process, such as milling depth and machine speed. Table~\ref{tab:full_summary_stats} provides descriptive statistics and variable definitions for all selected features included in the analysis. Continuous variables such as skid number and macrotexture depth exhibit substantial variability. For example, the measured skid number ranges from as low as 10.5 to as high as 47.0 across different time points, while macrotexture depth after milling spans from 0.48 to 3.58 mm.  Categorical variables were encoded numerically to support machine learning workflows. For instance, surface type is coded as 0 for HMA and 1 for seal coat, while climatic zone is simplified into two categories: dry–freeze (0) and dry–non-freeze (1), based on available data. Similarly, drum type is represented as 0 for fine milling drums (300 teeth) and 1 for standard drums (150 teeth). 

\begin{table}[H]
\centering
\footnotesize
\caption{Descriptions and summary statistics for all selected variables}
\label{tab:full_summary_stats}
\begin{tabular}{p{3.8cm} p{4.8cm} cccc}
\toprule
\textbf{Variable} & \textbf{Description / Encoding} & \textbf{Mean} & \textbf{Std.\ Dev.} & \textbf{Min} & \textbf{Max} \\
\midrule
Climatic Zone  & 0 = dry--freeze,\;1 = dry--non-freeze & 0.40 & 0.49 & 0 & 1 \\
Depth (In) & Milling depth in inches & 0.46 & 0.11 & 0.20 & 0.50 \\
Drum & 0 = Fine drum (300 teeth), 1 = Standard drum (150 teeth) & 0.70 & 0.46 & 0 & 1 \\
Macrotexture (mm) & Measured texture depth & 1.44 & 0.69 & 0.28 & 2.89 \\
Macrotexture after milling (mm) & Texture depth after milling & 2.37 & 0.59 & 0.48 & 3.58 \\
Macrotexture before milling (mm) & Texture depth before milling & 0.68 & 0.32 & 0.16 & 1.73 \\
Month & Months since treatment (0 = baseline) & 9.75 & 5.77 & 3.00 & 18.00 \\
Skid Number & Friction measure at given month & 29.91 & 9.59 & 10.50 & 47.00 \\
Skid Number after milling & Friction measure immediately after milling & 42.37 & 12.15 & 15.00 & 58.00 \\
Skid Number before milling & Friction measure before milling & 15.70 & 8.00 & 9.00 & 35.00 \\
Speed (fpm) & Milling machine speed (ft/min) & 68.52 & 17.81 & 30.00 & 100.00 \\
Surface Type & 0 = HMA,\;1 = Seal Coat & 0.59 & 0.50 & 0 & 1 \\
\bottomrule
\end{tabular}
\end{table}

\begin{figure}[H]
    \centering
    \includegraphics[width=0.9\linewidth]{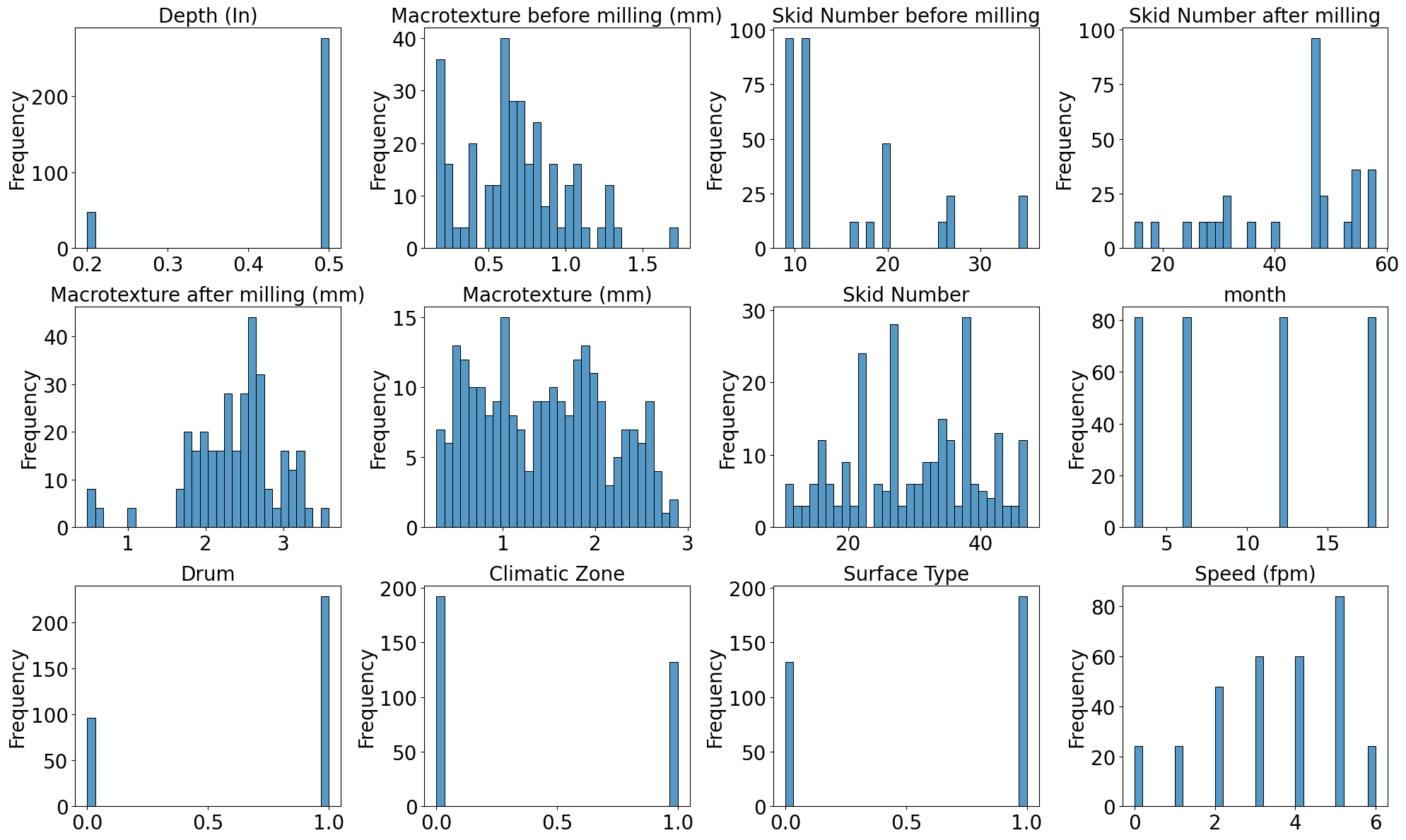}
    \caption{Histogram}
    \label{fig:histogram}
\end{figure}

Figure~\ref{fig:histogram} shows the distribution of all selected variables used in the modeling process. The skid number before milling is heavily left-skewed, which indicates that most segments started with relatively low friction levels. In contrast, skid number after milling shifts to higher values, which demonstrates the immediate effectiveness of surface treatment in improving friction. The skid number measured over time displays a more balanced distribution, capturing the degradation trend across months. The month variable itself shows a discrete distribution corresponding to scheduled inspection intervals (3, 6, 12, and 18 months). Categorical variables such as drum type, climatic zone, and surface type were encoded numerically and appear as bimodal distributions. These reflect the study design, where standard and fine drums, dry–freeze and dry–non-freeze regions, and HMA versus seal coat surfaces were all included in roughly balanced proportions. 

\begin{figure}[H]
    \centering
    \includegraphics[width=0.75\linewidth]{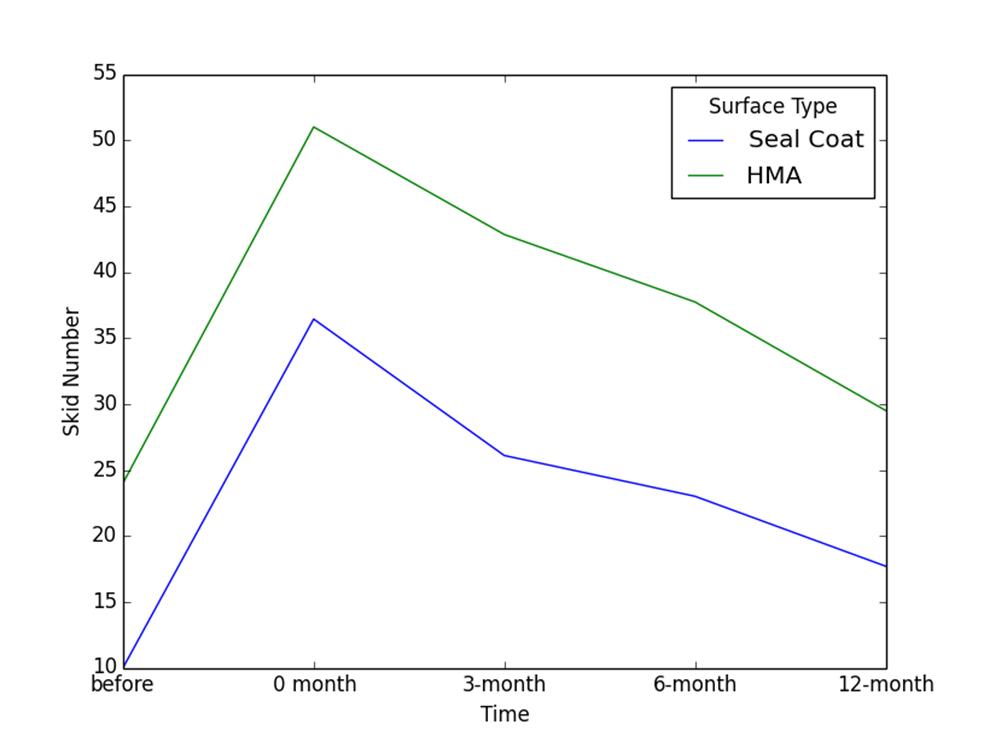}
    \caption{Skid number on different surface type}
    \label{fig:skid_surface}
\end{figure}

Figure \ref{fig:skid_surface} illustrates the evolution of skid number following micro-milling treatment on two different surface types: seal coat and HMA. Prior to treatment, both surfaces exhibited relatively low skid resistance—approximately 10 for seal coat and around 25 for HMA. Immediately after micro-milling (at 0 month), both surface types experienced a sharp increase in skid number, rising to about 37 for seal coat and as high as 51 for HMA. This demonstrates the effectiveness of micro-milling in restoring surface friction by increasing surface macrotexture. Notably, the improvement was more significant on HMA surfaces, likely due to their denser aggregate structure and better bonding with the milling process. 

However, starting from the third month, a clear downward trend in skid resistance is observed for both surface types. For seal coat, the skid number dropped rapidly, reaching around 26 at 3 months and declining to below 17 by 12 months, nearly reverting to pre-treatment levels. In contrast, HMA surfaces showed a more gradual reduction and maintained a skid number close to 30 at the 12-month mark. These results suggest that HMA surfaces offer better durability and resistance to polishing, which allows the benefits of micro-milling to persist longer than on seal coat pavements.

\begin{figure}[H]
    \centering
    \includegraphics[width=0.75\linewidth]{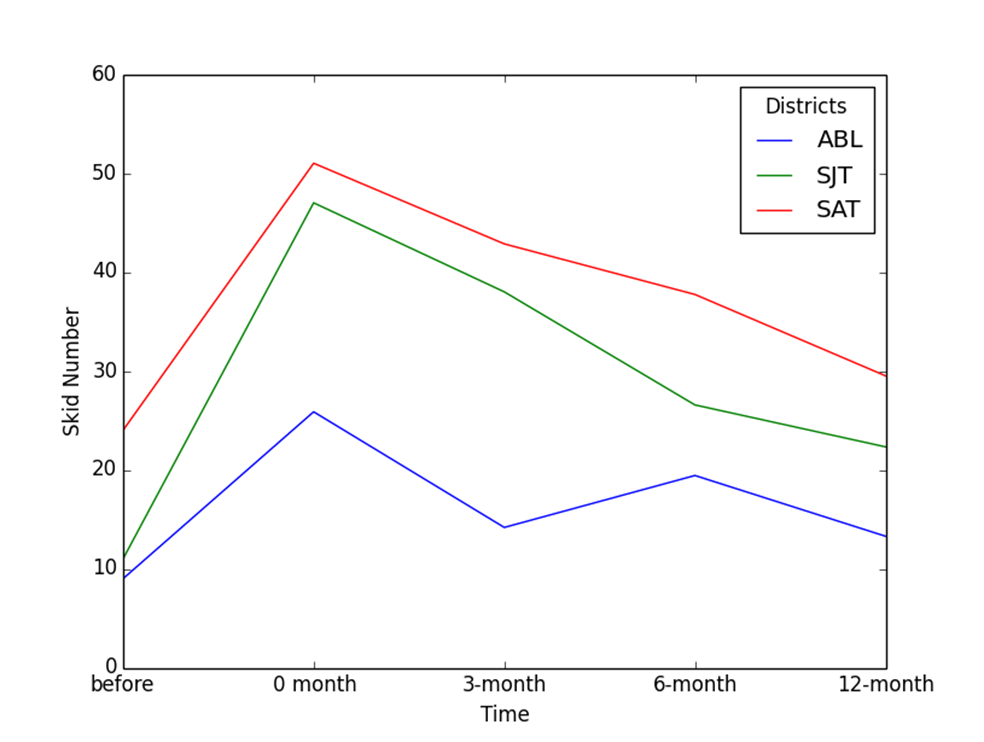}
    \caption{Skid number from different districts}
    \label{fig:skid_district}
\end{figure}

Figure \ref{fig:skid_district} compares the skid performance of selected sections from three different districts: San Antonio (SAT), San Angelo (SJT), and Abilene (ABL). All districts showed substantial improvement in skid number at 0 month, with SAT and SJT rising to 51 and 46, respectively, while ABL peaked at only 27. From “3 months” onward, district-level differences in deterioration rate became apparent. SAT maintained relatively high skid resistance over time, remaining above 30 at 12 months. In contrast, ABL experienced a rapid decline despite a brief rebound at 6 months, falling back to about 15 by the end of the observation period. 

These differences likely reflect the influence of local climate and traffic conditions. For example, SAT may have milder environmental exposure or lower truck traffic volumes, preserving surface texture for a longer duration. ABL, on the other hand, could be subject to harsher weather or more aggressive traffic loading, which accelerates wear and texture degradation.

\subsection{Modeling Results}

This section compares the performance of eight traditional machine–learning baselines with a time series transformer regression Model. The Transformer takes a sliding window of length~$L$ (here $L=4$) and predicts the skid number or macrotexture value at the next inspection month. For fair comparison, all models were evaluated on the same train–test split (80 \% / 20 \%) and assessed with the coefficient of determination ($R^2$), root-mean-squared error (RMSE), and mean-absolute error (MAE).

The time series transformer model is implemented in PyTorch and consists of two stacked encoder layers with four self-attention heads and a hidden dimension of 64. Input features at each time step are projected into a 64-dimensional latent space, followed by the addition of learnable positional encodings. The output from the last time step is passed through a two-layer feed-forward network with a hidden size of 32 and ReLU activation to produce a scalar prediction. The model is trained using the Adam optimizer with a learning rate of $10^{-3}$ and a batch size of 16, using mean squared error as the loss function. Table~\ref{tab:skid_results} shows that the Transformer achieves the highest accuracy for skid-number forecasting ($R^2=0.981$), edging out XGBoost ($R^2=0.979$) and outperforming all other baselines by a clear margin.  Tree-based models (Random Forest, Decision Tree) rank next, whereas linear regressions and the MLP network trail far behind, indicating that non-linear interactions and temporal dependencies are critical for capturing friction degradation.

\begin{table}[H]
\centering
\caption{Prediction results for Skid Number.}
\label{tab:skid_results}
\begin{tabular}{lccc}
\toprule
\textbf{Model} & \textbf{$R^2$} & \textbf{RMSE} & \textbf{MAE} \\
\midrule
Time series transformer & \textbf{0.981} & \textbf{1.42} & \textbf{0.84} \\
XGBoost & 0.979 & 1.46 & 0.87 \\
Random Forest & 0.967 & 1.81 & 1.08 \\
Decision Tree & 0.962 & 1.96 & 0.82 \\
k-Nearest Neighbors & 0.920 & 2.83 & 1.35 \\
Ridge Regression & 0.776 & 4.73 & 3.82 \\
Linear Regression & 0.775 & 4.74 & 3.88 \\
Lasso Regression & 0.726 & 5.23 & 4.42 \\
MLP Regressor & 0.641 & 5.99 & 4.77 \\
\bottomrule
\end{tabular}
\end{table}

The results of the macrotexture prediction are summarized in Table~\ref{tab:texture_results}. The Random Forest remains the top performer ($R^2=0.838$), but the Transformer closely follows with $R^2=0.831$ and nearly identical error metrics (RMSE = 0.28; MAE = 0.22).  
XGBoost ranks third, while linear models again underperform. These findings suggest that, although tree ensembles capture texture variability very well, the Transformer’s sequence modelling adds only marginally more benefit for this target.

\begin{table}[H]
\centering
\caption{Prediction results for Macrotexture.}
\label{tab:texture_results}
\begin{tabular}{lccc}
\toprule
\textbf{Model} & \textbf{$R^2$} & \textbf{RMSE} & \textbf{MAE} \\
\midrule
Random Forest & \textbf{0.838} & \textbf{0.27} & 0.22 \\
Time series Transformer & 0.831 & 0.28 & 0.22 \\
XGBoost & 0.809 & 0.30 & 0.22 \\
Decision Tree & 0.741 & 0.34 & 0.27 \\
k-Nearest Neighbors & 0.719 & 0.36 & 0.29 \\
Ridge Regression & 0.686 & 0.38 & 0.31 \\
Linear Regression & 0.685 & 0.38 & 0.31 \\
MLP Regressor & 0.646 & 0.40 & 0.33 \\
Lasso Regression & 0.494 & 0.48 & 0.40 \\
\bottomrule
\end{tabular}
\end{table}

Across both targets, ensemble trees (Random Forest, XGBoost) and the Transformer dominate, confirming that non-linear effects, feature interactions, and temporal patterns are essential for accurate forecasting. The Transformer’s slight advantage on skid number demonstrates the value of explicitly modeling sequential dependencies, whereas its comparable performance on macrotexture suggests diminishing returns when historical texture inherently exhibits weaker temporal autocorrelation. These results highlight the practicality of the time series transformer for proactive pavement-management applications, while also showing that well-tuned ensemble methods remain competitive baselines when computational simplicity is desired.

\section{Discussion}

This study investigated the deterioration of pavement skid resistance and macrotexture following micro-milling treatments, using a sequence-based Transformer model alongside traditional machine learning approaches. The results offer valuable insights into both the behavior of surface characteristics over time and the effectiveness of predictive modeling techniques.

The observed deterioration pattern, initial improvement in skid resistance due to binder removal, followed by a decline from aggregate polishing, and eventual stabilization, is consistent with previous studies. The sharper decline in seal coat sections compared to HMA aligns with findings by Wei et al. \citep{wei2022study}, who noted that surface type significantly influences the rate of friction loss. The superior durability of HMA surfaces observed in this study may be attributed to their denser aggregate structure and better resistance to polishing, as also reported by Zhang et al. \citep{zhang2024impact}.

The Transformer model outperformed all other models in predicting skid resistance (R² = 0.981), highlighting the importance of capturing temporal dependencies in pavement performance data. This supports recent advances in deep learning for infrastructure modeling, such as those by Liu et al. \citep{liu2022developing} and Gao et al. \citep{gao2015milled}, who demonstrated the value of sequence-aware models in predicting long-term pavement behavior. However, for macrotexture prediction, the Random Forest model slightly outperformed the Transformer, suggesting that macrotexture may be more influenced by static or non-temporal factors.

\subsubsection{Practical Implications and Limitations}

The ability to accurately forecast skid resistance degradation has direct implications for pavement management systems. By leveraging models like the Transformer, transportation agencies can proactively schedule maintenance, prioritize high-risk segments, and optimize resource allocation. Compared to traditional empirical models such as the Penn State or PIARC IFI models, which often lack flexibility and fail to account for environmental variability, the data-driven approach used here offers improved adaptability and predictive power.

Moreover, the inclusion of diverse variables, such as milling depth, drum type, and climatic zone, enhances the generalizability of the model across different operational contexts. This aligns with the recommendations of Koné et al. \citep{kone2023application}, who emphasized the importance of multi-dimensional input features in improving skid resistance prediction accuracy.

Despite its strengths, the study has several limitations. The dataset, while diverse, is limited in size and temporal scope, covering only up to 18 months post-treatment. This restricts the ability to model long-term deterioration trends or capture seasonal effects. Additionally, the study focused exclusively on micro-milling, limiting the generalizability of the findings to other preventive maintenance methods such as micro-surfacing or fog seals. Variability in construction practices and data quality across districts may also introduce unquantified noise into the models.

\subsubsection{Future Research}

Future work should aim to expand the dataset both temporally and geographically, incorporating multi-year observations and additional treatment types. Integrating structural indicators and cost data could enable life-cycle cost analysis and optimization, as suggested. Furthermore, the use of high-resolution sensing technologies and in-vehicle data could enhance model inputs and support real-time pavement condition monitoring. Finally, exploring interpretable AI techniques could help uncover the underlying mechanisms driving surface deterioration and improve stakeholder trust in model outputs.

\section{Conclusions}

\subsection{Summary of Findings}
This study systematically evaluated the short- to medium-term effects of milling treatments on pavement skid resistance and macrotexture. A comprehensive dataset from various climatic zones in Texas was used, including 324 samples containing construction parameters, environmental conditions, and time-series surface characteristics. A structured sliding window dataset was constructed to capture the temporal evolution of surface performance after treatment.

The results demonstrated that ensemble models such as XGBoost and Random Forest outperformed conventional methods like linear regression in predicting both skid number and macrotexture. More importantly, the proposed time series transformer-based modeling framework further improved predictive accuracy. Specifically, it achieved the highest $R^2 score =0.980$  for skid number prediction and also ranked second for macrotexture prediction, which validates the effectiveness of modeling surface deterioration as a dynamic sequence prediction task. These findings support the integration of temporal learning techniques into pavement performance modeling, which enables more proactive and data-driven maintenance strategies.

\subsection{Limitations}
This study focused exclusively on milling as the treatment method and was based on a relatively limited number of field samples with a constrained temporal range. Although the dataset includes a variety of construction, surface, and climate-related variables, it still suffers from data quality inconsistencies, missing values, and unquantified field-level variations such as construction management practices.

In addition, several important factors that influence pavement surface deterioration, including traffic volume, the percentage of heavy vehicles, seasonal variation, and construction timing, were not included in the dataset due to limited data availability. The absence of these variables may introduce bias into the predictive model, especially in relation to traffic loading and environmental conditions. Future studies should consider incorporating these factors to improve model robustness and generalizability.

\subsection{Recommendations for Future Work}
Several future research areas are recommended. (1) Future studies may incorporate additional preventive treatments such as micro-surfacing or fog seal to compare their long-term impacts on surface friction. (2) Expanding the time horizon to cover multi-year post-treatment performance would enhance the model’s practical relevance. (3) Integrating cost data and structural indicators could extend the framework to life-cycle cost analysis and maintenance optimization. (4) Incorporating high-resolution sensing data or in-vehicle measurements could enrich the input space and improve the model’s generalizability and real-time applicability.

\bibliographystyle{unsrtnat}
\bibliography{references}

\end{document}